\documentclass[aps,twocolumn,showpacs]{revtex4}
\usepackage{amsmath}
\usepackage{epsfig}

\begin{document}

\title{Looking for the hidden-charm pentaquark resonances in $J/\psi p$ scattering}
\newcommand*{\NJNU}{Department of Physics, Nanjing Normal University, Nanjing, Jiangsu 210023, China}\affiliation{\NJNU}

\author{Hongxia Huang}\email{hxhuang@njnu.edu.cn}\affiliation{\NJNU}
\author{Jun He}\email{junhe@njnu.edu.cn}\affiliation{\NJNU}
\author{Jialun Ping}\email{jlping@njnu.edu.cn}\affiliation{\NJNU}

\begin{abstract}
In the framework of quark delocalization color screening model, the three new reported pentaquarks 
$P_{c}(4312)$, $P_{c}(4440)$, and $P_{c}(4457)$ can be identified as the hidden-charm molecular
states $\Sigma_{c}D$ with $J^{P}=\frac{1}{2}^{-}$, $\Sigma_{c}D^{*}$ with $J^{P}=\frac{3}{2}^{-}$, and 
$\Sigma_{c}D^{*}$ with $J^{P}=\frac{1}{2}^{-}$, in the baryon-meson scattering process, respectively.
Besides, the $\Sigma^{*}_{c}D^{*}$ of both $J^{P}=\frac{1}{2}^{-}$ and $J^{P}=\frac{3}{2}^{-}$ are also 
possible molecular pentaquarks. Moreover, the calculation is extended to the $P_{c}-$like molecular pentaquarks 
$P_{b}$. Several states with masses above $11$ GeV and narrow width are obtained. All these heavy pentaquarks 
are worth searching in the future experiments.
\end{abstract}


\maketitle

\setcounter{totalnumber}{5}

\section{\label{sec:introduction}Introduction}
In 2015, the claim of two hidden-charm pentaquark states $P_{c}(4380)$ and
$P_{c}(4450)$ by the LHCb Collaboration~\cite{LHCb} attracted people's interesting in
the pentaquarks with heavy quarks and inspired a lot of theoretical work on these two states, such as the
baryon-meson molecules~\cite{ChenR1,ChenHX1,RocaL,HeJ,HuangHX,GangY,Meissner,XiaoCW2,ChenR2,ChenHX2,Azizi},
the diquark-triquark pentaquarks~\cite{Lebed,ZhuR},
the diquark-diquark-antiquark pentaquarks~\cite{Maiani,Anisovich,Ghosh,WangZG},
the genuine multiquark states~\cite{Mironov}, the topological soliton~\cite{Scoccola},
and the kinematical threshold effects in the triangle singularity mechanism~\cite{GuoFK,LiuXH1,Mikhasenko},
and so on. The lattice QCD simulation of $NJ/\psi$ and $N\eta_{c}$ scattering is also performed to find these 
$P_{c}$ states~\cite{Skerbis}.

Four years later, at the Rencontres de Moriond QCD Conference, the LHCb Collaboration reported the observation 
of three new pentaquarks, named as $P_{c}(4312)$, $P_{c}(4440)$, and $P_{c}(4457)$~\cite{Skw}. The $P_{c}(4312)$ 
was discovered with $7.3\sigma$ significance by analyzing the $J/\psi p$ invariant mass spectrum. The previously 
reported $P_{c}(4440)$ structure was resolved at $5.4\sigma$ significance into two narrow states: the $P_{c}(4440)$ 
and $P_{c}(4457)$. The masses and widths of these states are:
\begin{eqnarray}
P_{c}(4312): M & = & 4311.9\pm 0.7^{+6.8}_{-0.6}~ \mbox{MeV}, \nonumber\\
\Gamma & = & 9.8\pm 2.7^{+3.7}_{-4.5} ~\mbox{MeV},  \nonumber \\
P_{c}(4440): M & = & 4440.3\pm 1.3^{+4.1}_{-4.7}~ \mbox{MeV}, \nonumber\\
\Gamma & = & 20.6\pm 4.9^{+8.7}_{-10.1} ~\mbox{MeV},  \nonumber \\
P_{c}(4457): M & = & 4457.3\pm 0.6^{+4.1}_{-1.7}~ \mbox{MeV}, \nonumber\\
\Gamma & = & 6.4\pm 2.0^{+5.7}_{-1.9} ~\mbox{MeV}.
\end{eqnarray}
As mentioned in Ref.~\cite{Skw}, since all three states are narrow and below the $\Sigma^{+}_{c}\bar{D}^{0}$ and $\Sigma^{+}_{c}\bar{D}^{*0}$ thresholds within plausible hadron-hadron binding energies, they provide the 
strongest experimental evidence to date for the existence of molecular states composed of a charmed baryon and 
an anticharmed meson. Immediately after the report of the LHCb Collaboration, several theoretical work have been 
done to study the mass spectrum of these states~\cite{ChenHX3,ChenR3,LiuMZ}. Ref.~\cite{GuoFK2} studied the 
isospin breaking decays of the molecular structure of the $P_{c}(4457)$.

Searching for the existence of multiquark states is an important issue of the hadron physics. To provide the 
necessary information for experiments, mass spectrum calculation alone is not enough. The study of hadron-hadron 
scattering, as well as the main production process of multiquark states, is indispensable. In the framework of 
the quark delocalization color screening model (QDCSM), the detail of which can be found in Refs.~\cite{QDCSM0,QDCSM1}, 
we apply the well developed resonating group method (RGM)~\cite{RGM} to calculate the baryon-meson scattering phase 
shifts and to find the hidden-charm and hidden-bottom pentaquark resonances. The wave function of the baryon-meson 
system is of the form
\begin{equation}
\Psi = {\cal A } \left[\hat{\phi}_{A}(\boldsymbol{\xi}_{1},\boldsymbol{\xi}_{2})
       \hat{\phi}_{B}(\boldsymbol{\xi}_{3})\chi_{L}(\boldsymbol{R}_{AB})\right].
\end{equation}
where $\boldsymbol{\xi}_{1}$ and $\boldsymbol{\xi}_{2}$ are the internal coordinates for the baryon
cluster A, and $\boldsymbol{\xi}_{3}$ is the internal coordinate for the meson cluster B.
$\boldsymbol{R}_{AB} = \boldsymbol{R}_{A}-\boldsymbol{R}_{B}$ is the relative coordinate between
the two clusters. The $\hat{\phi}_{A}$ and $\hat{\phi}_{B}$ are the internal cluster wave functions of
the baryon A (antisymmetrized) and meson B, and $\chi_{L}(\boldsymbol{R}_{AB})$ is the relative motion
wave function between two clusters. The symbol ${\cal A }$ is the anti-symmetrization operator defined as
\begin{equation}
{\cal A } = 1-P_{14}-P_{24}-P_{34},
\end{equation}
where 1, 2, and 3 stand for the quarks in the baryon cluster and 4 stands for the quark in the meson cluster.
For a bound-state problem, $\chi_{L}(\boldsymbol{R}_{AB})$ is expanded by gaussian bases
\begin{eqnarray}
& & \chi_{L}(\boldsymbol{R}_{AB}) = \frac{1}{\sqrt{4\pi}}(\frac{6}{5\pi b^2})^{3/4} \sum_{i=1}^{n} C_{i}  \nonumber \\
&& ~~~~\times  \int \exp\left[-\frac{3}{5b^2}(\boldsymbol{R}_{AB}-\boldsymbol{S}_{i})^{2}\right] Y_{LM}(\hat{\boldsymbol{S}_{i}})d\hat{\boldsymbol{S}_{i}} \nonumber \\
&& ~~~~~~~~~~~~= \sum_{i=1}^{n} C_{i} \frac{u_{L}(R_{AB},S_{i})}{R_{AB}}Y_{LM}(\hat{\boldsymbol{R}}_{AB}).
~~~~~
\end{eqnarray}
with
\begin{eqnarray}
& & u_{L}(R_{AB},S_{i}) = \sqrt{4\pi}(\frac{6}{5\pi b^2})^{3/4}R_{AB}   \nonumber \\
&& ~~~~\times \exp\left[-\frac{3}{5b^2}(R^{2}_{AB}-S^{2}_{i})\right] i^{L} j_{L}(-i \frac{6}{5b^{2}}R_{AB}S_{i}).
~~~~~
\end{eqnarray}
where $\boldsymbol{S}_{i}$ is called the generating coordinate, $C_{i}$ is expansion coefficients,
$n$ is the number of the gaussian bases, which is determined by the stability of the results,
and $j_{L}$ is the $L$-th spherical Bessel function.

For a scattering problem, the relative wave function is expanded as
\begin{equation}
\chi_{L}(\boldsymbol{R}_{AB}) = \sum_{i=1}^{n} C_{i}
    \frac{\tilde{u}_{L}(R_{AB},S_{i})}{R_{AB}}Y_{LM}(\hat{\boldsymbol{R}}_{AB}) .
\end{equation}
with
\begin{eqnarray}
 & & \tilde{u}_{L}(R_{AB},S_{i}) =    \nonumber \\
 & & \left\{ \begin{array}{ll}
 \alpha_{i} u_{L}(R_{AB},S_{i}),   ~~~~~~~~~~~~~~~~~~~~~~~~~~~~~~~~~~R_{AB}\leq R_{C}  \\
  \left[h^{-}_{L}(k_{AB},R_{AB})-s_{i}h^{+}_{L}(k_{AB},R_{AB})\right] R_{AB},  R_{AB}\geq R_{C}
 \end{array} \right. \nonumber \\
\end{eqnarray}
where $h^{\pm}_{L}$ is the $L$-th spherical Hankel functions, $k_{AB}$ is the momentum of relative motion
with $k_{AB}=\sqrt{2\mu_{AB}E_{cm}}$, $\mu_{AB}$ is the reduced mass of two hadrons (A and B) of the open
channel; $E_{cm}$ is the incident energy, and $R_{C}$ is a cutoff radius beyond which all the strong
interaction can be disregarded. Besides, $\alpha_{i}$ and $s_{i}$ are complex parameters which are
determined by the smoothness condition at $R_{AB}=R_{C}$ and $C_{i}$ satisfy $\sum_{i=1}^{n} C_{i}=1$.
After performing variational procedure, a $L$-th partial-wave equation for the scattering problem can be
deduced as
\begin{eqnarray}
& & \sum_{j=1}^{n} {\cal L }^{L}_{ij} C_{j} = {\cal M }^{L}_{i}  ~~~~(i=0,1,\cdot\cdot\cdot, n-1), \label{Lij}
~~~~~
\end{eqnarray}
with
\begin{eqnarray}
& & {\cal L }^{L}_{ij} = {\cal K }^{L}_{ij}-{\cal K }^{L}_{i0}-{\cal K }^{L}_{0j}+{\cal K }^{L}_{00},
~~~~~
\end{eqnarray}
\begin{eqnarray}
& & {\cal M }^{L}_{i}  = {\cal K }^{L}_{00}-{\cal K }^{L}_{i0},
~~~~~
\end{eqnarray}
and
\begin{eqnarray}
& & {\cal K }^{L}_{ij}=\left\langle \hat{\phi}_{A}(\boldsymbol{\xi}'_{1},\boldsymbol{\xi}'_{2})\hat{\phi}_{B}(\boldsymbol{\xi}'_{3})
  \frac{\tilde{u}_{L}(R'_{AB},S_{i})}{R'_{AB}}Y_{LM}(\hat{\boldsymbol{R}}'_{AB}) \right.  \nonumber \\
& & ~~~~~~~~~\left|H-E\right| \nonumber \\
& & \left. {\cal A } \left[\hat{\phi}_{A}(\boldsymbol{\xi}_{1},\boldsymbol{\xi}_{2})\hat{\phi}_{B}(\boldsymbol{\xi}_{3})
  \frac{\tilde{u}_{L}(R_{AB},S_{j})}{R_{AB}}Y_{LM}(\hat{\boldsymbol{R}}_{AB})\right]\right\rangle. \nonumber \\
\end{eqnarray}
By solving Eq.(\ref{Lij}), we can obtain the expansion coefficients $C_{i}$. Then the $S$ matrix element $S_{L}$ and the phase shifts $\delta_{L}$ are given by
\begin{eqnarray}
& & S_{L} \equiv e^{2i\delta_{L}} = \sum_{i=1}^{n} C_{i}s_{i},
~~~~~
\end{eqnarray}

Resonances are unstable particles usually observed as bell-shaped structures in scattering cross sections of 
their corresponding open channels. For a simple narrow resonance, the peak position and the of the half-width of
the bell shape are the mass $M$ and the decay width $\Gamma$ of the resonance. The cross-section $\sigma_{L}$ and 
the scattering phase shifts $\delta_{L}$ have relations:
\begin{equation}
\sigma_{L} = \frac{4\pi}{k^{2}} (2L+1) \sin^2\delta_{L},  \label{sec}
\end{equation}
where $k=\sqrt{2\mu E_{cm}}/\hbar$; $\mu$ is the reduced mass of two hadrons of the open channel; $E_{cm}$ is the 
incident energy. Therefore, by using the scattering phase shifts, the cross sections can be easily obtained. 
The scattering phase shifts are shown in our previous work~\cite{HuangHX2}. Here, to compare with the experimental 
data, we calculate the cross sections of the $J/\psi p$ scattering channel, which are shown in Fig. 1. Based on our 
previous theoretical work, the new experimental information and the calculated cross sections, we discuss possible 
explanations of the three new pentaquarks: $P_{c}(4312)$, $P_{c}(4440)$, and $P_{c}(4457)$.

For $P_{c}(4312)$, our previous study showed that there was a narrow resonance state $\Sigma_{c}D$ with 
$J^{P}=\frac{1}{2}^{-}$ in the scattering channels of $\eta_{c}p$, $J/\psi p$, $\Lambda_{c}D$ and 
$\Lambda_{c}D^{*}$~\cite{HuangHX2}. The calculated mass of this resonance state is $4306.7\sim 4311.3$ MeV, 
and the decay width is $7.1$ MeV. From the Fig. 1(a), we also see a sharp peak appearing at the mass of 
$4307.9$ MeV with a very narrow partial width of about $1.2$ MeV. It is obvious that both the mass and decay width 
are close to the experimental value of the $P_{c}(4312)$, which indicates that the newly reported $P_{c}(4312)$ state 
can be identified as the $\Sigma_{c}D$ molecular pentaquark with $J^{P}=\frac{1}{2}^{-}$ in our model calculation.

\begin{figure}
\begin{center}
\epsfxsize=3.2in \epsfbox{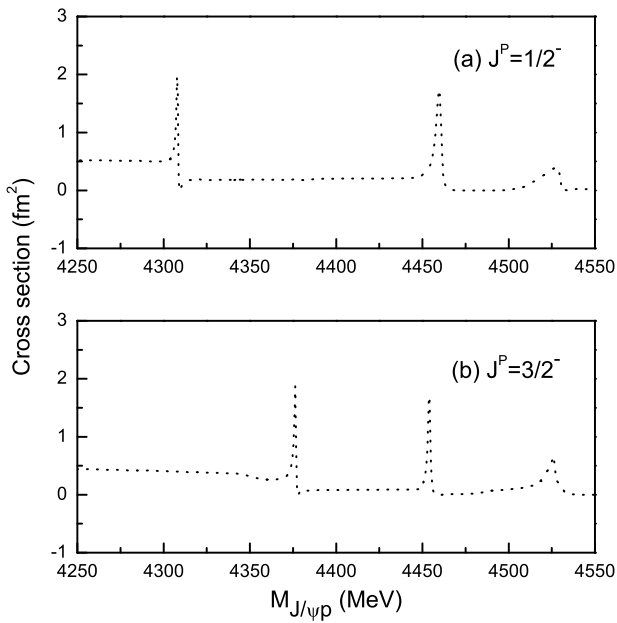} \vspace{+0.1in}

\caption{The cross section of the $J/\psi p$ channel with $J^{P}=\frac{1}{2}^{-}$ and $J^{P}=\frac{3}{2}^{-}$ respectively.}
\end{center}
\end{figure}

For $P_{c}(4440)$ and $P_{c}(4457)$, we assigned them as the resonances $\Sigma_{c}D^{*}$ with $J^{P}=\frac{1}{2}^{-}$ 
and $\Sigma_{c}D^{*}$ with $J^{P}=\frac{3}{2}^{-}$ according to our calculation~\cite{HuangHX2}. These two resonances 
also appear as two sharp peaks in the cross section of the $J/\psi p$ channel (see Fig. 1 (a) and (b)). The masses and
the partial decay widths can be read from Fig. 1, they are: $\Sigma_{c}D^{*}$ of $J^{P}=\frac{1}{2}^{-}$, $4459.7$ MeV 
and $3.9$ MeV; $\Sigma_{c}D^{*}$ of $J^{P}=\frac{3}{2}^{-}$, $4445.7$ MeV and $1.5$ MeV. Compared with experimental data, 
the $P_{c}(4440)$ is more possible
to be the molecular pentaquark $\Sigma_{c}D^{*}$ of $J^{P}=\frac{3}{2}^{-}$, and the $P_{c}(4457)$ can be explained as 
the molecular pentaquark $\Sigma_{c}D^{*}$ of $J^{P}=\frac{1}{2}^{-}$.

Besides the three resonances discussed above, one may see the fourth peak showed in Fig. 1 (b). It is the molecular 
pentaquark $\Sigma^{*}_{c}D$ with $J^{P}=\frac{3}{2}^{-}$. The mass of this resonance is $4376.4$ MeV, which is very 
close to the reported $P_{c}(4380)$. However, the decay width is only $1.5$ MeV, much smaller than the experimental 
value. We propose the experiment to find whether there is any narrow resonance near the $P_{c}(4380)$.

In addition, in Fig. 1(a), we also find there is a cusp near the mass of $4527$ MeV, which is the threshold of the 
$\Sigma^{*}_{c}D^{*}$. Our previous calculation showed that the single channel 
$\Sigma^{*}_{c}D^{*}$ of $J^{P}=\frac{1}{2}^{-}$ was bound, but it disappeared by coupling with the 
scattering channel $J/\psi p$, $\Sigma_{c}D$, and $\Sigma_{c}D^{*}$ channels~\cite{HuangHX2}. So it only appears as 
a cusp in the cross section of the $J/\psi p$. This is consistent with the experimental result, there is no any 
distinct signal near the threshold of the $\Sigma^{*}_{c}D^{*}$ in the experimental data. However, this molecular 
pentaquark $\Sigma^{*}_{c}D^{*}$ of $J^{P}=\frac{1}{2}^{-}$ appeared as a resonance state in the $\eta_{c} p$ 
scattering channel~\cite{HuangHX2}. The mass and width of it is $4525.8$ MeV and $4.0$ MeV, respectively. 
Besides, the $\Sigma^{*}_{c}D^{*}$ state of $J^{P}=\frac{3}{2}^{-}$ has the similar situation. It appears as a cusp 
in the cross section of the $J/\psi p$ (see Fig. 1(b)), but it turned to be a resonance state in the $\Lambda_{c}D^{*}$ 
scattering process, with a mass of $4523.0$ MeV and a width of $1.0$ MeV~\cite{HuangHX2}. Although it is difficult 
for the experiment to perform the $\eta_{c} p$ or the $\Lambda_{c}D^{*}$ scattering, the $\Sigma^{*}_{c}D^{*}$ of both $J^{P}=\frac{1}{2}^{-}$ and $J^{P}=\frac{3}{2}^{-}$ are possible molecular pentaquarks, which are worth looking for.

Because of the heavy flavor symmetry, we also extend the study to the hidden-bottom pentaquarks.
The results are similar to the hidden-charm molecular pentaquarks. Fig.2 shows the cross sections of the $\Upsilon p$ 
channel with $J^{P}=\frac{1}{2}^{-}$ and $J^{P}=\frac{3}{2}^{-}$ respectively.
From Fig. 2(a), we can see three sharp peaks in the cross sections, which correspond to three resonance states: 
$\Sigma_{b}B$, $\Sigma_{b}B^{*}$ and $\Sigma^{*}_{b}B^{*}$ states with $IJ^{P}=\frac{1}{2}\frac{1}{2}^{-}$. 
The resonance mass of $\Sigma_{b}B$ state is $11077.5$ MeV and the decay width is about $0.1$ MeV; 
$\Sigma_{b}B^{*}$ has the mass of $11125.8$ MeV and the decay width of $0.8$ MeV; and 
$\Sigma^{*}_{b}B^{*}$ has the mass of $11153.5$ MeV and the decay width of $3.0$ MeV.

\begin{figure}
\begin{center}
\epsfxsize=3.2in \epsfbox{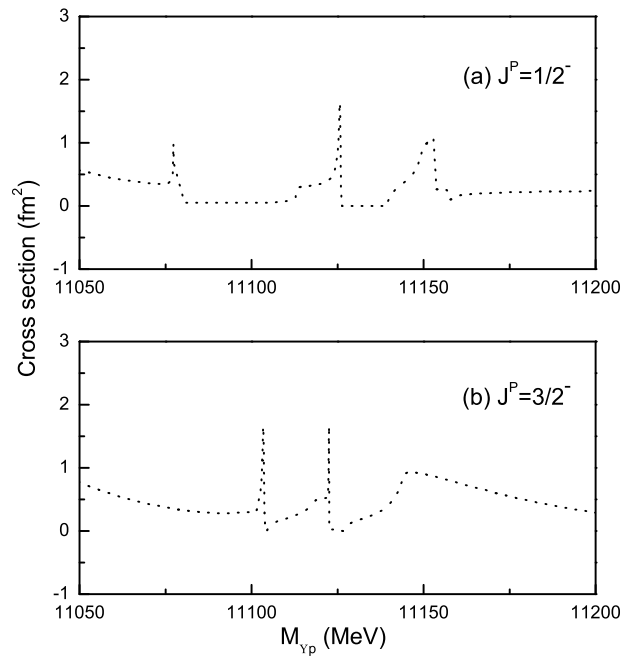} \vspace{+0.1in}

\caption{The cross section of the $\Upsilon p$ channel with $J^{P}=\frac{1}{2}^{-}$ and $J^{P}=\frac{3}{2}^{-}$ respectively.}
\end{center}
\end{figure}

For the hidden-bottom pentaquarks with $IJ^{P}=\frac{1}{2}\frac{3}{2}^{-}$, two pinnacles appear in the Fig. 2(b), 
corresponding to two resonance states: $\Sigma_{b}B^{*}$ and $\Sigma^{*}_{b}B$. The mass and the decay width of
$\Sigma_{b}B^{*}$ is $11122.7$ MeV and $0.2$ MeV respectively. The mass and the decay width of $\Sigma^{*}_{b}B$ 
is $11103.6$ MeV and $0.8$ MeV, respectively. All these hidden-bottom pentaquarks have the similar properties with 
the hidden-charm pentaquarks, so we can call them $P_{c}-$like molecular pentaquarks $P_{b}$, which are also worth 
searching for by experiments.

To summarize, in the framework of QDCSM, we look for the hidden-charm and hidden-bottom pentaquark resonances by 
studying $J/\psi p$ and $\Upsilon p$ scattering process. The three new narrow pentaquarks $P_{c}(4312)$, $P_{c}(4440)$, 
and $P_{c}(4457)$ observed in the process of $\Lambda_{b}\rightarrow J/\psi pK$ reported by LHCb can be interpreted 
as the hidden-charm molecular pentaquarks $\Sigma_{c}D$ with $J^{P}=\frac{1}{2}^{-}$, $\Sigma_{c}D^{*}$ with 
$J^{P}=\frac{3}{2}^{-}$, and $\Sigma_{c}D^{*}$ with $J^{P}=\frac{1}{2}^{-}$, respectively. Another molecular pentaquark 
$\Sigma^{*}_{c}D$ with $J^{P}=\frac{3}{2}^{-}$ is also existed in our calculation, the mass of which is close to the 
$P_{c}(4380)$, but the width is much smaller than it. Besides, the $\Sigma^{*}_{c}D^{*}$ of both $J^{P}=\frac{1}{2}^{-}$ 
and $J^{P}=\frac{3}{2}^{-}$ are possible molecular pentaquarks. All these narrow pentaquarks are worth searching for 
or being confirmed in future experiments. The Jefferson Lab has proposed to look for the hidden-charm pentaquarks by 
using photo-production of $J/\psi$ at threshold in Hall C~\cite{Jlab1}. Moreover, the pentaquarks with charm quarks 
can also be observed by the PANDA/FAIR~\cite{Panda}. For the pentaquarks with the hidden-bottom, we predict several 
$P_{c}-$like molecular pentaquarks $P_{b}$ above $11$ GeV with narrow width. We hope the proposed electron-ion 
collider (EIC)~\cite{EIC} and the upgraded facilities at Jefferson Lab~\cite{Jlab2} can play important role in
discovering these interesting super-heavy pentaquarks.

Searching for multiquark states is an important topic in hadron physics. To provide more information for
experiments, the baryon-meson scattering process calculation is expected. Doing baryon-meson scattering is
also a challenge for quark model. The model is needed to describe the baryon and meson spectra at the same time.
With the accumulation of more experimental data on baryon-meson scattering and the pentaquark, the quark model
will be further updated. To use the simple picture to describe the natural phenomena is one of the goal of physics.

\section*{Acknowledgment}
This work is supported partly by the National Science Foundation
of China under Contract Nos. 11675080, 11775118 and 11535005, the Natural Science Foundation of
the Jiangsu Higher Education Institutions of China (Grant No. 16KJB140006).

\end{document}